\documentclass[aps,prl,reprint,twocolumn,superscriptaddress]{revtex4-2}

\usepackage{amsmath} 
\usepackage{amssymb} 
\usepackage{graphicx}
\usepackage{dcolumn}
\usepackage{bm}
\usepackage[hidelinks]{hyperref}
\usepackage{color} 
\usepackage{siunitx}
\usepackage{comment}
\usepackage[capitalise]{cleveref}  
\crefname{section}{Sec.}{Secs.}%
\usepackage{soul,xcolor}
\usepackage[normalem]{ulem}

\usepackage[nearskip=0pt,farskip=0pt,captionskip=0pt,caption=false]{subfig}

\newcommand{\phantomsubfloat}[1]{{
    \captionsetup[subfigure]{labelformat=empty}
    ~\\[-1.6em]
    \subfloat[][]{#1}
}}

\crefrangelabelformat{figure}{#3#1#4--#5(\crefstripprefix{#1}{#2}#6}
\crefmultiformat{figure}{Figs.~#2#1\xdef\mycreffirstarg{#1}#3}{ and~#2({\crefstripprefix{\mycreffirstarg}{#1}}#3}{, #2({\crefstripprefix{\mycreffirstarg}{#1}}#3}{ and~#2({\crefstripprefix{\mycreffirstarg}{#1}}#3}

\newcommand{\blue}[1]{{\color{blue} #1}}

\newcommand{\Fig}[2][]{\cref{fig:#2}\if #1\empty\else(#1)\fi}

\newcommand\LDst{\bgroup\markoverwith{\blue{\rule[0.5ex]{2pt}{0.4pt}}}\ULon}

\newcommand{\kb}{k_\text{B}}  
\renewcommand{\exp}[1]{\text{exp}\left[{#1}\right]} 

\newcommand{\affilTracy}{Institut f\"ur Experimentalphysik, Universit\"at Innsbruck, Technikerstrasse 25, 6020 Innsbruck, Austria}
\newcommand{\affilPhotonics}{Photonics Laboratory, ETH Z\"{u}rich, 8093 Z\"{u}rich, Switzerland}
\newcommand{\affilQC}{Quantum Center, ETH Z\"urich, Z\"urich, Switzerland}
\newcommand{\affilOriolone}{Institute for Theoretical Physics, University of Innsbruck, Innsbruck, Austria}
\newcommand{\affilOrioltwo}{Institute for Quantum Optics and Quantum Information,
Austrian Academy of Sciences, Innsbruck, Austria}

\begin{document}

\title{State Expansion of a Levitated Nanoparticle in a Dark Harmonic Potential}

\author{Eric \surname{Bonvin}}
\author{Louisiane \surname{Devaud}}
\author{Massimiliano \surname{Rossi}}
\author{Andrei \surname{Militaru}}
\affiliation{\affilPhotonics}
\affiliation{\affilQC}

\author{Lorenzo \surname{Dania}}
\affiliation{\affilTracy}
\author{Dmitry~S.~\surname{Bykov}}
\affiliation{\affilTracy}

\author{Oriol \surname{Romero-Isart}}
\affiliation{\affilOriolone}
\affiliation{\affilOrioltwo}
\author{Tracy E. \surname{Northup}}
\affiliation{\affilTracy}

\author{Lukas \surname{Novotny}}
\author{Martin \surname{Frimmer}}
\affiliation{\affilPhotonics}
\affiliation{\affilQC}

\date{\today}


\begin{abstract}

Levitated nanoparticles in vacuum are prime candidates for generating macroscopic quantum superposition states of massive objects. Most protocols for preparing these states necessitate coherent expansion beyond the scale of the zero-point motion to produce sufficiently delocalized and pure phase-space distributions. Here, we spatially expand and subsequently recontract the thermal state of a levitated nanoparticle by modifying the stiffness of the trap holding the particle. We achieve state-expansion factors of 25 in standard deviation for a particle initially feedback-cooled to a center-of-mass thermal state of \SI{155}{\milli\kelvin}. Our method relies on a hybrid scheme combining an optical trap, for cooling and measuring the particle's motion, with a Paul trap for expanding its state. Consequently, state expansion occurs devoid of measurement backaction from photon recoil, making this approach suitable for coherent wavefunction expansion in future experiments. 
 
\end{abstract}

\maketitle

\paragraph*{Introduction.}

Quantum physics predicts that a massive object can be prepared in a macroscopic quantum superposition state---a state in which the center-of-mass position is delocalized over length scales larger than the object's size~\cite{arndt2014testing}. Matter-wave experiments have demonstrated this remarkable phenomenon with macromolecules as large as $10^4$ atomic mass units~\cite{fein2019quantum}. Over the past decade, levitated nanoparticles in vacuum~\cite{millen2019optomechanics, gonzalez-ballestero2021levitodynamics} have emerged as promising candidates to increase the mass of such macroscopic quantum states by at least four orders of magnitude~\cite{romero-isart2011large, romero-isart2011quantum, scala2013matterwave, bateman2014nearfield, wan2016free, pino2018onchip, stickler2018probing, neumeier2022fast, roda-llordes2023macroscopic}. 
The first important merit of these levitated systems is that their motional state can be purified by ground-state cooling~\cite{delic2020cooling, magrini2021realtime, tebbenjohanns2021quantum, kamba2022optical, Ranfagni2022, piotrowski2023simultaneous}.
Their second important merit is that levitation offers the opportunity to  manipulate the trapping potential and thereby control the motional state of a levitated object~\cite{rashid2016experimental, hebestreit2018sensing, frimmer2019rapid, ciampini2021experimental}.
This second merit sets levitated systems apart from clamped optomechanical oscillators~\cite{aspelmeyer2014cavity}. In those systems, impressive optomechanical quantum control has been achieved~\cite{Rossi2018, oconnell2010quantum, lecocq2015quantum, riedinger2016nonclassical, moller2017quantum, chu2018creation, sletten2019resolving, bild2023schrodinger}, but the length scale set by the zero-point motion of the oscillator cannot be tuned in situ.

Control over the trapping potential is a key requirement in both approaches pursued to expand the wave-function of a levitated nanoparticle. The first approach relies on purely optical potentials together with free evolution in the absence of a potential~\cite{neumeier2022fast}. 
The second approach aims to exploit evolution in a dedicated potential in the absence of decoherence due to measurement back-action inevitably associated with optical interactions~\cite{jain2016direct, maurer2023quantum, dania2023ultrahigh}. 
To this end, this strategy proposes to use dark, non-optical potentials generated by radio-frequency (RF) electric fields~\cite{roda-llordes2023macroscopic,pino2018onchip}.
On the experimental side, hybrid approaches of optical traps combined with RF (Paul) traps have been investigated~\cite{millen2015cavity, nagornykh2015cooling, fonseca2016nonlinear, delord2017diamonds, conangla2018motion, conangla2020extending, bykov2022hybrid} but remain to be exploited for expansion protocols. 

Generating and handling a quantum mechanically pure state of a nanoparticle in a hybrid trap is a daunting experimental challenge. Nevertheless, expansion protocols can already be tested on classical states to accelerate future progress when moving to the quantum regime.

In this work, we expand the classical thermal state of motion of a levitated nanoparticle by transferring it from an optical trap to a weakly confining dark potential implemented with a linear Paul trap.
After evolution in the dark potential, the particle is re-trapped optically and its position is measured, allowing for repetition of the protocol.
We demonstrate a maximum state expansion by a factor of 25 in standard deviation and subsequent contraction to the starting size in addition to a contribution from heating during the protocol.
This work is an important step towards coherent wavefunction expansion of levitated nanoparticles in the quantum regime.

\begin{figure}[b]
    \includegraphics[width=1\columnwidth]{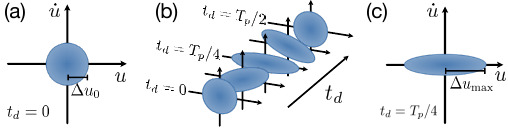}
    \phantomsubfloat{\label{fig:phase_space_theory_init}}
    \phantomsubfloat{\label{fig:phase_space_theory_evo}}
    \phantomsubfloat{\label{fig:phase_space_theory_final}}
    \caption{
    Phase-space representation of a particle in a harmonic potential undergoing a frequency-jump protocol.
    (a)~The particle is initially in a thermal state, described by a uniform Gaussian distribution of size $\Delta u_0$. 
    (b)~The state evolves in the weaker potential for increasing $t_d$.
    (c)~The maximum size $\Delta u_\text{max}$ is reached for $t_d=T_p/4$.
    }        
    \label{fig:phase_space_theory}
\end{figure}

\paragraph*{Frequency-jump protocol.}
We start by illustrating the dynamics of a harmonic oscillator with position $u$ and velocity $\dot{u}$ under a so-called frequency-jump protocol~ \cite{janszky1986squeezing, graham1987squeezing, lo1990squeezing, cosco2021enhanced, Xin2021}.
In this protocol, we initialize the oscillator in a thermal state 
in a (stiff) potential with frequency $\omega_o$ and period $T_o$ (in this work implemented with an optical trap).
With appropriately normalized axes, this initial thermal state is described in phase space by a Gaussian distribution with widths $\Delta u_0$ and $\Delta\dot u_0$ along the position and velocity axes, respectively, schematically shown in \cref{fig:phase_space_theory_init}.
At time $t=0$, we switch to a weaker potential with frequency $\omega_p<\omega_o$ and period $T_p$ (implemented with a Paul trap in the absence of laser light) for a time $t_d$.
The evolution of the state in the weaker potential as a function of $t_d$ is illustrated in Fig.~\ref{fig:phase_space_theory_evo}. 
As $t_d$ increases, the state expands spatially until reaching a maximum size $\Delta u_\text{max}$ for $t_d=T_p/4$, as shown in \cref{fig:phase_space_theory_final}. As $t_d$ increases further, the spatial state size returns to its initial value, reached at  $t_d=T_p/2$.
To characterize the phase-space distribution acquired at time $t_d$, we instantaneously switch the potential back to frequency $\omega_o$. Now, the state rotates in phase space with preserved shape. In this work, we experimentally realize this frequency-jump protocol in a hybrid Paul-optical trap.

\begin{figure}[t]
    \includegraphics[width=1\columnwidth]{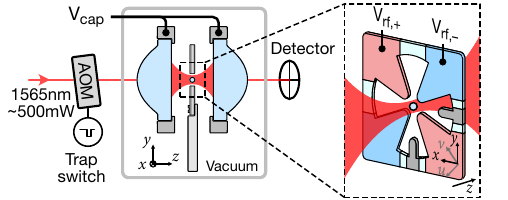}
    \caption{
    Experimental setup. A laser beam is focused by a lens to form an optical trap. 
    The trapped particle's center-of-mass motion is recorded on a quadrant photodetector.
    The laser can be switched on and off with an acousto-optic modulator (AOM).
    The optical trap is superimposed with a Paul trap. It consists of a microfabricated chip held in the focal plane and endcap electrodes (held at a potential $V_\mathrm{cap}$). 
    The two pairs of on-chip electrodes are driven with RF voltages ($V_\mathrm{rf,+}$ and $V_\mathrm{rf,-}$, respectively) for in-plane confinement. 
    }
    \label{fig:setup}
\end{figure}

\paragraph*{Experimental setup.}
Our setup overlaps an optical trap with a linear Paul trap, in which we levitate a \SI{177}{\nano\meter} diameter silica nanosphere. This configuration, shown in \cref{fig:setup}, allows us to selectively trap the same particle in either trap. What follows is a simplified description of the setup  (further details  in Ref.~\cite{Bonvin2023}).

The optical trap is generated by a laser ($x$-polarized, wavelength \SI{1565}{\nano\meter}, power \SI{500}{\milli\watt}) focused by a lens ($\text{NA}=0.77$, optical axis along $z$). The particle's center-of-mass oscillation frequencies are $2\pi\times$(44,\,58,\,10)~\SI{}{\kilo
\hertz} along $(x,y,z)$, respectively. The trap can be switched on and off with an acousto-optic modulator on a timescale faster than $100$~ns. 
The forward scattered light is collected by an identical lens, and sent onto a quadrant photodiode for position detection in the $(x,y,z)$-basis.

The Paul trap consists of a glass substrate micromachined to a wheel-trap geometry (FEMTOprint, Switzerland) \cite{chen2017ticking, chen2017sympathetic, brewer201927, teller2023integrating}. The metallized electrodes are oriented along the orthogonal directions $u$ and $v$, which lie in the focal plane and are tilted at $45^\circ$ relative to $x$ and $y$. The two pairs of RF electrodes are driven with opposite phases at a frequency of \SI{33}{\kilo\hertz} and a nominal peak-to-peak voltage of \SI{400}{\volt}. Confinement along $z$ is achieved by applying a typical endcap voltage of \SI{70}{\volt} to the metallic lens holders. This results in secular center-of-mass modes at $2\pi\times$(6,\,6,\,3)~\SI{}{\kilo\hertz} along $(u,v,z)$, respectively. 
The radial and axial mode frequencies can be tuned from \SIrange{0.1}{6}{\kilo\hertz} by adjusting the applied voltages. 
The trap is mounted in a vacuum chamber and kept at a pressure of \SI{2.2e-6}{\milli\bar}.
Before starting our experiments, we compensate for stray fields in the trapping region using the shim electrodes of our trap~\cite{Bonvin2023,BonvinThesis}.

\begin{figure}[t]
    \includegraphics[width=1\columnwidth]{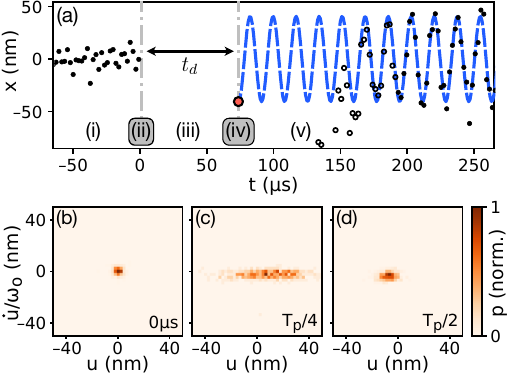}
    \phantomsubfloat{\label{fig:protocol}}
    \phantomsubfloat{\label{fig:phase_space_exp_1}}
    \phantomsubfloat{\label{fig:phase_space_exp_2}}
    \phantomsubfloat{\label{fig:phase_space_exp_3}}
    \caption{
    (a)~Recorded position during a single realization of a frequency-jump protocol (black data points). The optical trap is disabled from $t=0$ to $t=t_d=\SI{72}{\micro\second}$, allowing the particle to evolve in the Paul trap. 
    The oscillation after recapture is fit (dashed blue line) to estimate the position and velocity of the particle at time of recapture $t=t_d$ (red circle).
    (b)~
    Initial state of the particle, reconstructed from 500 repetitions of the protocol with $t_d=0$ (bin size $\SI{1}{\nano\meter}\times\SI{1}{\nano\meter}$).
    (c)~For $t_d=\SI{72}{\micro\second}\approx T_p/4$, the state has visibly expanded along the position axis.
    (d)~For $t_d=\SI{140}{\micro\second}\approx T_p/2$, the state has contracted back to a size similar to the initial distribution.
    }
    \label{fig:protocol_and_phase_space_exp}
\end{figure}

\paragraph*{Results.}
We execute a frequency-jump protocol by rapidly transferring the particle from the high-frequency optical trap to the lower frequency Paul trap and back.
Experimentally, this is achieved by switching the optical trap off and back on, while keeping the Paul trap enabled at all times.
More specifically, we execute the following protocol:
(i)~for $t<0$, the particle is held in the optical trap, and initialized to a feedback-cooled thermal state of motion;
(ii)~at time $t=0$, feedback cooling is disabled, and the optical trap is rapidly switched off;
(iii)~for $0<t<t_d$, the particle evolves in the Paul trap until
(iv)~at $t=t_d$, the optical trap is switched back on to recapture the particle optically; and
(v)~the final position of the particle in phase space is characterized for $t>t_d$.
In \cref{fig:protocol}, we show the measured particle position during one realization of this frequency-jump protocol with $t_d = \SI{72}{\micro\second}$ (black data-points). 
The measurement of the particle's position for times $t<0$ is largely impacted by noise, as expected from a state under feedback cooling. In contrast, the oscillations in the particle's position are clearly visible at the end of the protocol ($t>\SI{150}{\micro\second}$). 
We note that for times $0<t<t_d$, no measurement record exists, since the optical field is off. Furthermore, between $t=72~\mu$s and $\approx150~\mu$s,  we observe a slow recovery of the recorded time trace after re-enabling the optical trap due to AC-coupling of the data acquisition system.
To avoid unwanted artefacts in our data analysis, we ignore the first \SI{120}{\micro\second} of signal upon recapture [open data points in \cref{fig:protocol}, see~\cite{supplemental}].

We use the measurement record to retrodict the position and velocity after recapture at $t=t_d$ in the optical trap (see Supplemental Material~\cite{supplemental}). The retrodicted trajectory is shown as the dashed blue line in \cref{fig:protocol}.
The phase-space distribution before the frequency jump is inferred by retrodicting the phase-space position of the particle for protocols executed with $t_d=0$. Thus, all phase-space distributions are generated using the identical data-processing protocol.
Another challenge arises from the coupling of the motion along $x$ and $y$ by the electric fields in the Paul trap, which are aligned along the $u$ and $v$ axes. Therefore, in the following, we combine the measurements of $x$ and $y$ to analyze motion along $u$.

A phase-space distribution of the particle before the frequency jump, i.e., at $t_d=0$, generated from 500 repetitions of the experiment is shown in \cref{fig:phase_space_exp_1}. Each repetition is started at the same time in the micromotion cycle.
Here, the particle is cooled to a state size of \SI{1.5\pm0.1}{\nano\meter} in the optical trap. This corresponds to a center-of-mass temperature of $T_\text{CoM}=\SI{155\pm25}{\milli\kelvin}$. The state size $\Delta u$ is defined as the standard deviation of the position $u$. We characterize the measurement noise independently and subtract it from the signal when evaluating $\Delta u$.
The error is evaluated using the bootstrap method~\cite{efron1979bootstrap}.
\Cref{fig:phase_space_exp_2} shows the distribution after an evolution time $t_d=\SI{72}{\micro\second}$, corresponding to a quarter oscillation period in the Paul trap.
In comparison to the initial state, the distribution for $t_d=\SI{72}{\micro\second}$ is stretched along the position axis, reaching a state size of \SI{26.4\pm1.2}{\nano\meter}. This state expansion in a dark potential is the main result of this Letter.
In \cref{fig:phase_space_exp_3}, we show the distribution for $t=140~\mu s\approx T_p/2$, where the state has contracted to $\Delta u=\SI{4.4\pm0.2}{\nano\meter}$, as expected.
The increase in state size during evolution from \cref{fig:phase_space_exp_1} to \cref{fig:phase_space_exp_3} is a signature of heating due to collisions of the particle with gas molecules~\cite{jain2016direct, tebbenjohanns2019cold, conangla2019optimal}.

\begin{figure}[t]
    \includegraphics[width=1\columnwidth]{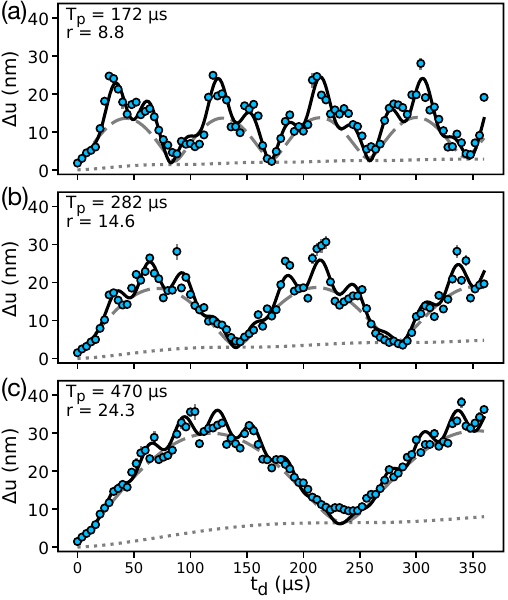}
    \phantomsubfloat{\label{fig:variance_stiff}}
    \phantomsubfloat{\label{fig:variance_medium}}
    \phantomsubfloat{\label{fig:variance_weak}}
    \caption{
    State size $\Delta u$ as a function of evolution time $t_d$, for different Paul trap stiffnesses, quantified by the frequency ratio $r$. 
    Solid lines: full model described in the Supplemental Material~\cite{supplemental}. 
    Dashed lines: simplified model discussed in the main text. 
    Dotted lines: heating contribution $\Delta u_h$ under the simplified model. 
    }
    \label{fig:variances_experiment}
\end{figure}

We proceed by investigating the role of the frequency ratio $r=\omega_o/\omega_p$ (tuned via the voltage of the Paul trap RF drive) in our expansion protocol.
\Cref{fig:variance_stiff} shows the measured state size $\Delta u$ (blue circles) as a function of evolution time $t_d$, for a frequency ratio $r=8.8$. 
We identify the following three features.
First, the state size grows and shrinks with a period $T_p/2$, as expected.
Second, we observe a fast modulation whose period matches that of the RF-field driving the Paul trap.
Third, the minimum of each expansion cycle trends upwards, suggesting that the minimal state size continuously grows as a consequence of heating during the expansion protocol.

We investigate these features for larger frequency ratios $r=14.5$ and $r=24.3$ in~\cref{fig:variance_medium,fig:variance_weak}, respectively. 
For weaker Paul traps, the main oscillations are slower and of larger amplitude than for stiffer traps. In~\cref{fig:variance_weak}, we achieve a state expansion by a factor 25.
We observe that weaker Paul traps are less heavily modulated by oscillations at the RF frequency, which we attribute to the lower RF drive voltages. Finally, weaker traps are more sensitive to heating, seen by the more rapid increase of the minima in~\cref{fig:variance_weak} as  compared to~\cref{fig:variance_stiff}.

\paragraph*{Discussion.}
To understand our observations, we model the behavior of a particle's state subject to a frequency-jump protocol. To simplify the description, we assume that the Paul trap is a harmonic potential. A complete model including the effects of micromotion can be found in Ref.~\cite{supplemental}.
We model the particle in the Paul trap as a harmonic oscillator of mass $m$ coupled to a thermal bath at rate $\gamma$, following the equation of motion
\begin{equation}
    \ddot{u} + \omega_p^2u = \sqrt{2\gamma k_\text{B} T/m}~\xi_\text{th},
    \label{eq:eom}
\end{equation}
where we neglect the damping term (since $\gamma\ll\omega_p$), we assume $u^2\ll{k_BT/(m\omega_p^2)}$, and each dot represents a time derivative. 
The bath has temperature $T$ and is modeled using the stochastic variable $\xi_\text{th}$, which satisfies $\left\langle\xi_\text{th}(t)\xi_\text{th}(t')\right\rangle = \delta(t-t')$. Here, $\left\langle\dots\right\rangle$ denotes an ensemble average. Solving \cref{eq:eom}, we find that the spatial state size of the particle, given by the standard deviation $\Delta u$ after the protocol, evolves according to $\Delta u^2(t) =\Delta u_c^2(t) + \Delta u_h^2(t)$. 
The first contribution stems from coherent evolution and is given by~\cite{graham1987squeezing}
\begin{equation}
    \Delta u_c^2(t) =
	\Delta u_0^2
    \cos^2 (\omega_p t) +
	r^2 \frac{
    \Delta \dot{u}_0^2
    }{\omega_o^2} \sin^2 (\omega_p t),
    \label{eq:coherent}
\end{equation}
with $\Delta u_0$ the initial state size. 
The second term of Eq.~\eqref{eq:coherent} scales with the ratio of the initial velocity state size $\Delta\dot u_0^2$ and the (square of the) frequency of the Paul trap $\omega_p^2=\omega_o^2/r^2$. When the protocol is started with a thermal state of the optical potential, we have $\Delta\dot u_0/\omega_o=\Delta u_0$ and the maximum spatial state size from coherent evolution is given by $\Delta u_{c,\text{max}}=r\Delta u_0$, such that the spatial expansion is simply dictated by $r$.
The situation in our experiment is slightly more complicated. The particle is initialized in the optical trap in two uncorrelated thermal states along $x$ and $y$. However, we perform our subsequent experiments in the rotated basis $uv$. The distributions along $u$ and $v$ then become correlated, and the relation $\Delta\dot u_0/\omega_o=\Delta u_0$ no longer holds. For more detail, refer to the supplemental material~\cite{supplemental}.
The second contribution to $\Delta u^2$ arises from heating and reads
\begin{equation}
    \label{eq:heating}
    \Delta u_h^2(t) =
	r^2\frac{\hbar\Gamma}{m\omega_o}
	\left(t - \frac{\sin (2\omega_p t)}{2\omega_p} \right).
\end{equation}
Here, to align with the literature~\cite{roda-llordes2023macroscopic}, we have introduced the heating rate $\Gamma$, which is expressed in units of phonons of the optical trapping potential. In our experiment, where heating is dominated by coupling to the residual gas in the vacuum chamber at temperature $T$, we have $\Gamma=\gamma k_B T/(\hbar\omega_o)$, which we measure to be $\Gamma=2\pi\times\SI{926\pm56}{\kilo\hertz}$.

Care must be taken when defining the optical trap frequency $\omega_o$ in Eq.~\eqref{eq:coherent}. Since we are interested in the particle motion along $u$, but the optical trap's eigenmodes are oriented along $x$, $y$, and $z$, we define the effective oscillation frequency along $u$ as $\omega_{u\text{,opt}}={[(\omega_{x\text{,opt}}^2 + \omega_{y\text{,opt}}^2)/2}]^{1/2}$. We fit our model to the measured data using the Paul trap frequency $\omega_p$ as the only fit parameter. 
All other parameters ($\Delta u_0$, $\Delta \dot{u}_0$, $\omega_o$, $\gamma$, and $T$) are measured independently.

The simplified model is shown in \cref{fig:variances_experiment} as the dashed gray lines, while the dotted gray lines show the heating contribution $\Delta u_h$. Our simplified model reproduces all features of the data well, except the modulation at the micromotion frequency.
In contrast, the full model (solid line in \cref{fig:variances_experiment}) accounts for all features observed during a frequency-jump protocol. 
Comparing the two models in~\cref{fig:variance_stiff} reveals that, for stiff Paul traps, the presence of the RF fields almost doubles the expanded state size as compared to a simple frequency jump.

Another important observation is that our simple model explains the increase of the heating contribution with increasing frequency ratio $r$, experimentally observed in~\cref{fig:variances_experiment}. 
Specifically, according to~\cref{eq:heating}, at compression time $T_p/2$, the total state size has no contribution due to coherent evolution, and is given by
\begin{equation}
    \label{eq:compression_time_state_size}
    \Delta u^2(T_p/2) = \Delta u_0^2 + r^2\frac{\hbar\Gamma}{m\omega_o} \frac{T_p}{2}.
\end{equation}
The only contribution to $\Delta u^2$ besides the initial value $\Delta u_0^2$ is from heating, with a rate amplified by $r^2$. 

Regarding the current limitations of our system,  larger expansions could, in principle, be achieved with the help of weaker potentials. 
However, the associated increase in sensitivity to stray fields leads to frequent failures to recapture the particle in the optical trap and must be overcome with more accurate stray field and gravity compensation in the future.

\paragraph* {Conclusion.}

We have constructed a hybrid Paul-optical trap with high-numerical-aperture optical access, as used in systems achieving measurement-based ground-state cooling of optically levitated nanoparticles~\cite{tebbenjohanns2021quantum,magrini2021realtime}. Using this platform, we have exploited the large frequency difference between optical and Paul traps to expand and subsequently recontract the thermal state size of an optically levitated nanoparticle by a factor of 25.
The expansion happens in the absence of photon recoil, which is critical for extending this protocol to generate macroscopic quantum states.

Having demonstrated expansion of a thermal state, a logical next step is to move to ultra-high vacuum, unlocking access to ground-state cooling~\cite{magrini2021realtime, tebbenjohanns2021quantum,kamba2022optical}.
In the quantum regime, an exciting prospect is to observe quantum squeezing of mechanical motion in the quadrature orthogonal to the one that is coherently expanded. Indeed, in the absence of decoherence an expansion factor of 25 would generate 28~dB of motional squeezing in variance.
A challenge in this endeavor is the increased sensitivity to heating during the protocol.
On the other hand, this increased sensitivity can potentially be exploited. Since state expansion in the Paul trap happens in the absence of laser light, our system may provide access to weak decoherence rates, such as from blackbody emission, which are overwhelmed by photon recoil in optically levitated systems \cite{romero-isart2011quantum,bateman2014nearfield,agrenius2023interaction}.
Finally, our platform could also be used to generate inverted quadratic potentials, by reversing the endcap voltages. This may provide an avenue to reach large expansions in short times~\cite{romero-isart2017coherent, weiss2021large,roda-llordes2023macroscopic}.

\begin{acknowledgments}
This research was supported by the Swiss National
Science Foundation (SNF) through Grant No.\ 212599 and No.\ 217122, NCCR-QSIT program (Grant No.\ 51NF40-160591), the European
Union’s Horizon 2020 research and innovation program
under Grant No.\ 951234 (Q-Xtreme), and by the Austrian Science Fund (FWF) Project No. Y951 and Grant No. I5540. 
\end{acknowledgments}

\bibliography{Free_falls_paper,Martin}





\clearpage
\onecolumngrid

\renewcommand{\thefigure}{S\arabic{figure}}
\renewcommand{\theequation}{S\arabic{equation}}
\renewcommand{\thesection}{S\arabic{section}}
\setcounter{equation}{0}
\setcounter{figure}{0}
\setcounter{section}{0}
\newcommand{\Supps}[1]{Supp.~\ref{sec:#1}}

\begin{center}
  \LARGE
  \textbf{Supplemental Materials}
\end{center}

\section{Change of basis}
\label{sec:SM_change_of_basis}
In this section, we analyze the state-space distribution of a harmonic oscillator under a coordinate transformation. Our goal is to find a relation between the state size in the coordinate system $uv$ of the Paul trap and that in the $xy$ system of the optical trap.

We consider two orthogonal modes $x$ and $y$ of a particle trapped in a harmonic potential. We express the velocities of the particle as $\dot{x}$ and $\dot{y}$, and the modes have resonance frequencies $\omega_x$ and $\omega_y$, respectively.
The particle is undergoing linear feedback cooling along both axes, such that we consider two effective temperatures $T_x$ and $T_y$.
Each mode is then in a thermal state, with uncorrelated positions and velocities, and the phase space probability distribution $\rho$ is given by
\begin{equation}
\label{eq: compact initial thermal state}
\rho(x, y, \dot{x}, \dot{y}) = \rho_q(x, y) \rho_p(\dot{x}, \dot{y}),
\end{equation}
with
\begin{subequations}
\label{eq: compact elements}
\begin{align}
\rho_q(x, y) &= \frac{m\omega_x \omega_y}{2\pi \kb \sqrt{T_x T_y}} \exp{- m\left( \frac{\omega_x^2 x^2}{2\kb T_x} + \frac{\omega_y^2 y^2}{2\kb T_y} \right)}, \\
\rho_p(\dot{x}, \dot{y}) &= \frac{m}{2\pi \kb \sqrt{T_x T_y}} \exp{- m\left( \frac{\dot{x}^2}{2\kb T_x} + \frac{\dot{y}^2}{2\kb T_y} \right)},
\end{align}
\end{subequations}
where we have separated the total probability distribution into its position contribution $\rho_q(x, y)$ and velocity contribution $\rho_p(\dot{x}, \dot{y})$. Both distributions in Eqs.~\eqref{eq: compact elements} have diagonal covariance matrices, which can be seen from the lack of cross terms.

We now introduce two orthogonal directions $u$ and $v$ which lie in the $xy$-plane and are tilted at $45^\circ$ with respect to $x$ and $y$. We perform a change of basis from $xy$ to $uv$ by applying
\begin{equation}
\label{eq: change of reference}
x = \frac{u + v}{\sqrt{2}}, \qquad
y = \frac{u - v}{\sqrt{2}}, \qquad
\dot{x} = \frac{\dot{u} + \dot{v}}{\sqrt{2}}, \qquad
\dot{y} = \frac{\dot{u} - \dot{v}}{\sqrt{2}}.
\end{equation}
We can perform this transformation on $\rho_q(x, y)$ and 
$\rho_p(\dot{x}, \dot{y})$ independently. We have
\begin{subequations}
\label{eq: both transformations}

\begin{equation}
\label{eq: position transformation}
\begin{aligned}
\tilde{\rho}_q(u, v) &= \frac{m\omega_x \omega_y}{2\pi \kb \sqrt{T_x T_y}} \exp{ -\frac{1}{2} \frac{m}{2\kb} \begin{pmatrix} {u+v} & {u-v} \end{pmatrix}  \begin{pmatrix} \omega_x^2/T_x & 0 \\ 0 & \omega_y^2/T_y \end{pmatrix} \begin{pmatrix} {u+v} \\ {u-v} \end{pmatrix} } \\
&= \frac{m\omega_x \omega_y}{2\pi \kb \sqrt{T_x T_y}} \exp{ -\frac{1}{2} \begin{pmatrix} u & v \end{pmatrix}  \underbrace{\frac{m}{2\kb} \begin{pmatrix} \omega_x^2/T_x + \omega_y^2/T_y & \omega_x^2/T_x - \omega_y^2/T_y \\ \omega_x^2/T_x - \omega_y^2/T_y & \omega_x^2/T_x + \omega_y^2/T_y \end{pmatrix}}_{\Sigma_q^{-1}} \begin{pmatrix} u \\ v \end{pmatrix} },
\end{aligned}
\end{equation}
where we explicitly highlight the covariance matrix $\Sigma_q$ in the new basis. With analogous steps, we derive 
\begin{equation}
\label{eq: momentum transformation}
\tilde{\rho}_p(\dot{u}, \dot{v}) = 
\frac{m}{2\pi \kb \sqrt{T_x T_y}} \exp{ -\frac{1}{2} \begin{pmatrix} \dot{u} & \dot{v} \end{pmatrix}  \underbrace{\frac{m}{2\kb} \begin{pmatrix} 1/T_x + 1/T_y & 1/T_x - 1/T_y \\ 1/T_x - 1/T_y & 1/T_x + 1/T_y \end{pmatrix}}_{\Sigma_p^{-1}} \begin{pmatrix} \dot{u} \\ \dot{v} \end{pmatrix} },
\end{equation}
\end{subequations}
from which the analogous definition of the new covariance matrix $\Sigma_p$ for the velocities follows. Finally, we find the expressions for the new covariance matrices:
\begin{subequations}
\label{eq: cov matrices}
\begin{align}
\Sigma_q &= \frac{\kb}{2m} \begin{pmatrix}
T_x/\omega_x^2 + T_y/\omega_y^2 && T_x/\omega_x^2 - T_y/\omega_y^2 \\
T_x/\omega_x^2 - T_y/\omega_y^2 && T_x/\omega_x^2 + T_y/\omega_y^2 \end{pmatrix},
\\
\Sigma_p &= \frac{\kb}{2m} \begin{pmatrix}
T_x + T_y && T_x - T_y \\
T_x - T_y && T_x + T_y \end{pmatrix}.
\end{align}
\end{subequations}
We can appreciate from Eqs.~\eqref{eq: both transformations} that the original Gaussian distribution remains Gaussian in the new frame of reference. However, we now observe correlations between $u$ and $v$, as well as between $\dot{u}$ and $\dot{v}$, whose magnitude depends on the difference between the temperatures and resonance frequencies. Interestingly, if the original axes have different resonance frequencies $\omega_x$ and $\omega_y$, it is not possible to set all cross correlation terms to zero simultaneously in Eqs.~\eqref{eq: cov matrices}: either the new positions or the new velocities will remain correlated. The position and velocity variances in the new basis can directly be read off from the covariance matrix:
\begin{equation}
    \label{eq:variances_uv_basis}
    \Delta u^2 = \Delta v^2 =
    \frac{\kb}{2m} \left(\frac{T_x}{\omega_x^2} +
    \frac{T_y}{\omega_y^2} \right),
    \qquad
    \Delta \dot{u}^2 = \Delta \dot{v}^2 =
    \frac{\kb}{2m} \left( {T_x} + {T_y} \right).
\end{equation}
From these relations, we conclude that the initial state does not satisfy $\Delta\dot{u}_0/\omega_o = \Delta u_0$. Instead, for our system parameters, we find $\Delta\dot{u}_0/\omega_o \simeq 0.85\Delta u_0$, indicating that the maximum achievable state size is $\Delta u_{c,\text{max}}=0.85r\Delta u_0$.

\section{Position and velocity measurement}
\label{sec:SM_data_analysis}
\paragraph*{Detector calibration.}
The particle's position is recorded in the form of electric signals generated by a quadrant photo-diode (QPD). We convert these signals into length units for a more meaningful quantitative analysis by following the process outlined in~\cite{hauer2013general,hebestreit2018calibration}. A measurement record of the particle in the optical trap is acquired at higher pressures (typically \SI{5}{\milli\bar}), under which conditions the particle is in thermal equilibrium with a bath at ambient temperature. This knowledge allows us to convert the measured signal (in volts) to the expected particle displacement (in meters). Calibration is performed separately for the $x$ and $y$ modes of the particle's motion.

\paragraph*{Processing of frequency-jump protocol measurement records.}
Next, let us describe the data analysis process used to extract the position and velocity of a particle after a frequency-jump protocol.
Each measurement record consists of two time traces, measuring the particle position along $x$ and $y$, respectively.
For each individual time trace, we perform the following three steps. First we truncate the time trace, only keeping the data \SI{120}{\micro\second} after particle recapture, i.e., during phase~(v) of the protocol. The \SI{100}{\micro\second} buffer serves to reject effects related to the AC-coupling of our data acquisition system from the data analysis. Second, we demodulate this data at the optical trap frequency. Third, we adjust the phase of the demodulated components to account for timing delays in the system. The in-phase and quadrature components of the final demodulated signal describe the particle's position and velocity upon recapture, respectively.

The filter stage of the demodulation process has a \SI{2}{\kilo\hertz} bandwidth and is applied backwards in time, effectively retrodicting the signal. Special care must be taken when choosing the demodulator frequency. Across a series of experiments, the laser power, and hence the trap frequency, is subject to slight fluctuations. For each realization, we adjust the demodulator frequency to precisely match that of the measured signal, preventing the filtering stage from introducing unwanted phase shifts. 

\begin{figure}[h]
    \centering
    \includegraphics[width=7cm]{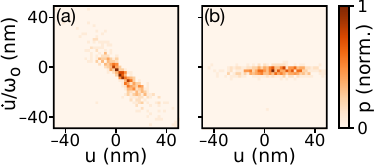}
    \caption{Example phase space distribution (a)~before and (b)~after delay adjustment. The variable trigger delay causes the phase space distribution to fan out, while the fixed delay adds an angle to the distribution.}
    \label{fig:SM_delay_adjust}
\end{figure}

Any timing delays during the measurement sequence manifest themselves as angled phase space distributions. These have several different origins. First, the trigger that launches the frequency-jump protocol is not precisely aligned to the sampling grid of the data acquisition system. This introduces a delay of \SIrange{0}{2}{\micro\second}, which we record individually for each iteration. Next, the response time of the AOM is limited by the velocity of the acoustic wave in the AOM crystal. This introduces a fixed delay measured to be \SI{0.6}{\micro\second}. Finally, we observe an additional delay of \SI{0.9}{\micro\second} of unknown origin. An example of the same phase space before and after accounting for timing delays is shown in~\cref{fig:SM_delay_adjust}. The variable trigger delay causes the phase space distribution to fan out, while the fixed delay adds an angle to the distribution.

\paragraph*{Characterization of measurement noise.}
In order to characterize the measurement noise of our system, we perform an identical data analysis process to the one described above, with the only difference being a shifted spectral window during the demodulation phase. Instead of demodulating our measured signals in a \SI{2}{\kilo\hertz} band around the frequency of our mode of interest, we shift this window to a spectral region with no signal. This resulting phase space distribution is the product of noise in our system. We refer to the position variance of this distribution as our measurement noise, which we measure to be \SI{2.5}{\nano\meter^2}. This noise is then subtracted from the measured variance of our particle during a frequency-jump protocol.

\section{Frequency-jump protocol in a Paul trap}
\label{sec:SM_full_model}

In this section, we describe the state size of a particle undergoing a frequency-jump protocol in a Paul trap. The derivation below closely follows \cite{joos1989langevin}. 
We assume a particle of mass $m$, with $n_e$ elementary charges, in a Paul trap. It is coupled with a rate $\gamma$ to a bath at temperature $T$. Neglecting friction, we assume that it follows the Mathieu equation \cite{mclachlan1964theory}
\begin{equation}
	\frac{d^2u}{d\tau^2}
	+ \left(a_u-2q_u\cos(2\tau)\right)u
	=  \sqrt{\frac{2\gamma k_\text{B} T}{m}}~\frac{4\xi_\text{th}(\tau)}{\Omega^2}
	\label{eq:eom_paul_trap}
\end{equation}
where $\Omega$ is the angular frequency of the RF driving tone, and we have introduced the dimensionless time variable $\tau=\Omega t/2$. Here, $a_u$ and $q_u$ are the conventional Mathieu parameters along $u$, given by
\begin{equation}
    a_u = \frac{4n_ee}{m\Omega^2}\frac{\partial^2\phi_\text{dc}}{\partial u^2} , \quad
    q_u = \frac{2n_ee}{m\Omega^2}\frac{\partial^2\phi_\text{rf}}{\partial u^2},
\end{equation}
with $\phi_\text{dc}$ and $\phi_\text{rf}$ the DC and RF potentials of the Paul trap, respectively.
Floquet theory \cite{meixner1954mathieu, abramowitz1964handbook} provides the following linearly independent solutions to the homogeneous equation of \cref{eq:eom_paul_trap}
\begin{equation}
  \begin{aligned}
  	\lambda_1(\tau)
  	 &= \sum_{n=-\infty}^{\infty}C_{2n}\cos((2n+\beta)\tau),\\
  	\lambda_2(\tau) &= \sum_{n=-\infty}^{\infty}C_{2n}\sin((2n+\beta)\tau),\\
  \end{aligned}
  \label{eq:homogeneous_soln}
\end{equation}
with $\beta$ the characteristic exponent of the Mathieu equation. From this, we can describe the full solution to the equation of motion \cref{eq:eom_paul_trap}
\begin{equation}
 u(\tau) = \tilde{u}_0\lambda_1(\tau)
 		   + (\tilde{v}_{u,0}/{\omega_p})\lambda_2(\tau)
 		   +\int_0^{\tau} g(\tau, \tau')
 		   \sqrt{\frac{2\gamma k_\text{B} T}{m}}~\frac{4\xi_\text{th}(\tau')}{\Omega^2}d\tau'
\label{eq:eom_solution_paul_trap}
\end{equation}
where $\tilde{u}_0$ and $\tilde{v}_{u,0}$ are related to the initial position and velocity by 
\begin{equation}
    u_0 = \tilde{u}_0\sum_n C_{2n} , \quad
    v_{u,0} = \tilde{v}_{u,0}\sum_n (2n+\beta) C_{2n}/\beta,
\end{equation}
and $\omega_p = \beta\Omega/2$ is the Paul trap's secular frequency. The Green's function $g(\tau, \tau')$ has the form
\begin{equation}
    g(\tau, \tau') = \frac{\left(\lambda_1(\tau')\lambda_2(\tau) - \lambda_1(\tau)\lambda_2(\tau')\right)}{W}
\end{equation}
where $W=\lambda_1\lambda'_2 - \lambda'_1\lambda_2$ is the Wronskian of the system. Substituting \cref{eq:homogeneous_soln}, we get
\begin{equation}
  g(\tau, \tau') =
  \frac{\sum_{n, n'}C_{2n}C_{2n'}
  \sin\left((2n+\beta)\tau -(2n'+\beta)\tau' \right)}{\sum_n(2n+\beta)C_{2n}^2}.
\end{equation}
We are now in a position to evaluate the state size of a particle evolving in a Paul trap for a duration $t$ using \cref{eq:eom_solution_paul_trap}. We break it down into its coherent and heating contributions according to $\Delta u^2(t) =\Delta u_c^2(t) + \Delta u_h^2(t)$. The contribution from coherent evolution is given by
\begin{equation}
    \Delta u_c^2(t) =
    \Delta\tilde{u}_0^2 \lambda^2_1(t)
 	+ r^2\frac{\Delta\tilde{u}_0^2}{\omega_\text{o}^2} \lambda_2^2(t)
    \label{eq:coherent_paul}
\end{equation}
and the contribution from heating is described by
\begin{equation}
    \Delta u_h^2(t) =
    r^2\frac{2\gamma k_\text{B}T}{m\omega_o^2}
 	\int_0^t g^2(t, t')dt'
    \label{eq:heating_paul}
\end{equation}
where we have introduced the frequency ratio $r=\omega_o/\omega_p$.

Both the coherent and heating contributions resemble their counterparts in the simplified model \cref{eq:coherent,eq:heating}, with the main difference being that the trigonometric functions are replaced with infinite sums of trigonometric functions, capturing the effect of micromotion. For the heating contribution \cref{eq:heating_paul}, an added consequence is that the integral of the Green's function can no longer be analytically evaluated. If we choose to apply the pseudo-potential approximation to \cref{eq:coherent_paul,eq:heating_paul} by keeping only the lowest order term, i.e., taking $C_{2n}=\delta_{n0}$, we recover \cref{eq:coherent,eq:heating}.
The full model used in the main text is numerically evaluated by keeping terms up to $n=\pm1$, using the expression $C_{\pm2}={-q_u}/{(4\pm4\beta)}$~\cite{abramowitz1964handbook}.

\end{document}